\documentstyle[11pt,aaspp4]{article}

\begin{document}
\title{Young Low-Mass Stars and Brown Dwarfs in IC~348}
\author{K. L. Luhman\altaffilmark{1}}
\affil{Harvard-Smithsonian Center for Astrophysics, 60 Garden St., Cambridge, 
MA 02138; kluhman@cfa.harvard.edu}

\altaffiltext{1}{Visiting Astronomer, Kitt Peak National Observatory, 
National Optical Astronomy Observatories, which is operated by the 
Association of Universities for Research in Astronomy, Inc.\ (AURA) under
cooperative agreement with the National Science Foundation.}

\begin{abstract}

I present new results from a continuing program to identify and 
characterize the low-mass stellar and substellar populations in 
the young cluster IC~348 (1-10~Myr). Optical spectroscopy has revealed
young objects with spectral types as late as M8.25. 
The intrinsic $J-H$ and $H-K$ colors of these sources are dwarf-like, 
whereas the $R-I$ and $I-J$ colors appear intermediate between the
colors of dwarfs and giants.
Furthermore, the spectra from 6500 to 9500~\AA\ are 
reproduced well with averages of standard dwarf and giant spectra, suggesting
that such averages should be used in the classification of young late-type
sources.

An H-R diagram is constructed for the low-mass population 
in IC~348 (K6-M8). The presumably coeval components of the young quadruple 
system GG~Tau (White et al.)\ and the locus of stars in IC~348 are 
used as empirical isochrones to test the theoretical evolutionary models. 
The calculations of Burrows et al.\ do not appear to be consistent with the 
data at these earliest stages of stellar evolution.  
There is fair agreement between the data and the model isochrones of 
D'Antona \& Mazzitelli, except near the hydrogen burning limit. The agreement
cannot be improved by changing the conversion between spectral types and
effective temperatures. 
On the other hand, for the models of Baraffe et al., an adjustment of the 
temperature scale to progressively warmer temperatures at later M types, 
intermediate between dwarfs and giants,  
brings all components of GG~Tau onto the same model isochrone and gives
the population of IC~348 a constant age and age spread as a function of mass.
When other observational constraints
are considered, such as the dynamical masses of GM~Aur, DM~Tau,
and GG~Tau~A, the models of Baraffe et al.\ are the most consistent 
with observations of young systems.  With compatible temperature scales,
the models of both D'Antona \& Mazzitelli and Baraffe et al.\ suggest
that the hydrogen burning mass limit occurs near M6 at ages of 
$\lesssim10$~Myr. Thus, several likely brown dwarfs are discovered
in this study of IC~348, with masses down to $\sim20$-30~$M_J$. 

\end{abstract}


\section{Introduction}

Brown dwarfs have been discovered over a wide range 
of ages.  Examples of evolved field brown dwarfs ($\gtrsim1$~Gyr) 
include the companion GL~229B (Nakajima et al.\ 1995; Oppenheimer et al.\ 1995)
and the free floating Kelu~1 (Ruiz, Leggett, \& Allard 1997), 
in addition to substellar objects identified through the near-infrared (IR) 
surveys of 2MASS (Kirkpatrick et al.\ 1999) and DENIS (Mart{\'\i}n et al.\ 1997;
Tinney, Delfosse, \& Forveille 1997; Delfosse et al.\ 1997). 
Warmer, more luminous brown dwarfs have been found at younger ages, such as
the field companion G~196-3B ($\sim300$~Myr; Rebolo et al.\ 1998) and
objects in the Pleiades (125~Myr) (Stauffer, Hamilton, \& Probst 1994;
Mart{\'\i}n et al.\ 1998, references therein) and in the youngest clusters 
($\lesssim10$~Myr) in Orion (Hillenbrand 1997),
$\sigma$~Orionis (B\'{e}jar, Zapatero Osorio, \& Rebolo 1999),
Taurus (Brice\~{n}o et al.\ 1998), $\rho$~Oph (Luhman, Liebert, \& Rieke 1997, 
hereafter LLR; Wilking, Greene, \& Meyer 1999), IC~348 (Luhman et al.\ 1998b, 
hereafter LRLL), Chamaeleon~I (Comer\'{o}n, Rieke, \& Neuh\"{a}user 1999),
and TW~Hydrae (Lowrance et al.\ 1999).  In a sample of candidate low-mass 
members of the Hyades, Reid \& Hawley (1999) have also serendipitously
discovered five likely pre-main-sequence (PMS) brown dwarfs 
possibly behind the Hyades and associated with the Taurus-Auriga star forming 
region. Given the relatively small numbers of these objects, it is not clear
whether they are representative of brown dwarfs at a given age, mass, and 
environment. Larger samples of brown dwarfs are critical for 
understanding the formation and evolution of substellar objects.

In this study, I provide a more complete picture of brown dwarfs and 
low-mass stars at their earliest stages of evolution ($<10$~Myr).
Work along these lines began with observations of V410~X-ray~3 in the
L1495E region of Taurus (Luhman et al.\ 1998a), which is one of the first M6
objects (0.06-0.1~$M_\odot$) discovered in a star forming region 
(Strom \& Strom 1994). The colors from $V$ through $L\arcmin$ were 
approximately dwarf-like with no significant near-IR excess emission.
The IR and optical spectra exhibited features indicative of both 
dwarfs and giants, a behavior that has been seen in the other 
young cool objects discovered subsequently (e.g., LRLL). Given its 
youth (1-10~Myr), proximity (300~pc), and rich, compact nature
(300 stars, $D\sim20\arcmin$), IC~348 is an excellent site for expanding 
this work to larger numbers and later types.  Recent studies of IC~348 
include IR imaging (Lada \& Lada 1995), optical
photometry (Trullols \& Jordi 1997), H$\alpha$ measurements
and optical images and spectroscopy (Herbig 1998),
and deeper IR photometry and optical and IR spectroscopy 
concentrated towards the $5\arcmin\times5\arcmin$ cluster core (LRLL).

I have begun to increase both the area and depth of 
the photometry and spectroscopy, with the goal of systematically identifying 
and characterizing the stellar and substellar populations within the entire 
cluster of IC~348. When this survey is completed, the analysis of LRLL
concerning the initial mass function (IMF), star formation history, and disk 
properties will be updated with better number statistics and 
completeness to lower substellar masses. However, with data recently collected, 
the known low-mass population has grown enough that the typical 
characteristics of young low-mass stars and brown dwarfs can be investigated. 

I will describe new optical imaging and spectroscopy of low-mass candidates,
discuss the optical classification of young late-type objects in
comparison to field dwarfs and giants, and present spectral types for the 
coolest sources (M4-M8.25). I will then examine the behavior of the $JHK$ colors
relative to dwarfs and to warmer, more massive ($\sim$M0, 0.5-1~$M_\odot$) 
classical T~Tauri stars (CTTS). After estimating reddenings, effective 
temperatures, and bolometric 
luminosities and constructing a Hertzsprung-Russell (H-R) diagram, the locus 
of objects in IC~348 and the presumably coeval components of the multiple
system GG~Tau (White et al.\ 1999, hereafter W99) will be used as empirical 
isochrones to test the theoretical evolutionary models and temperature scales.
With these results, I discuss the likely masses and ages of the objects 
observed in IC~348 and suggest that the hydrogen burning limit at young ages 
occurs near M6.

\section{Observations}
\label{sec:obs}

Optical images of IC~348 were obtained with the four shooter camera at the
Fred Lawrence Whipple Observatory 1.2~m telescope on 1998 September 23 under
photometric conditions.  The instrument contained four $2048\times2048$ 
CCDs at a plate scale of $0\farcs33$~pixel$^{-1}$, with the detectors separated 
by $\sim1\arcmin$ and arranged in a $2\times2$ grid.
Four positions were observed towards the center of IC~348 in dithers
of a few arcminutes, covering a total area of $25\arcmin\times25\arcmin$.
At each position, images were obtained at $R$ and $I$ with exposure times 
of 20~min.  The images were bias subtracted, 
divided by dome flats, registered, and combined into one image at each 
band. Image coordinates and photometry were extracted with DAOFIND and PHOT 
under the IRAF package APPHOT. The plate solution was derived from 
coordinates of all sources observed by Lada \& Lada (1995) that appeared in
the optical images and were not saturated.  
Saturation occurred near $I=16.5$ and the average 
completeness limits were $R\sim23$ and $I\sim21$, with brighter limits towards
the nebulosity in the cluster center. For instance, with the enhanced 
background, source 613 has a limit of $R>21.5$. Measurements were 
also hampered near the bright B stars BD+$31\arcdeg$643 and $o$~Per.
A few sources detected in the IR such as 611 fell within diffraction spikes 
of brighter stars and could not be measured.

The $R$-band filter in the four shooter camera has
the same shape as Cousins $R$ but with a full width at half maximum that
is 150~\AA\ smaller.  The $I$-band sensitivity of the four shooter 
is broader in wavelength than Cousins $I$, with a long tail to the red. 
Whereas Cousins $I$ falls from 90 to 5\% of the peak sensitivity from 
8500 to 9000~\AA, the four shooter maintains 80 and 60\% at these 
wavelengths, with 35 and 15\% at 9500~\AA\ and 1~\micron.
Standard stars calibrated in Cousins $R$ and $I$ by Landolt (1992) were
observed at colors of $R-I<1.2$ and $R-I=2.7$. Alternately, the 
transformation from the instrumental system to Cousins was modeled by
convolving spectra of M dwarfs at various types with the
instrument sensitivities -- detector quantum efficiencies and filter 
transmissions -- for the four shooter and the Cousins system (Landolt 1992). 
The color transformation derived in this modeling
agreed with the one measured from the photometry of the standards.  
I then modeled the effect of reddening on the uniqueness of the color 
correction, i.e. do a reddened M4 star and an unreddened M7 star of the same 
instrumental $R-I$ have the same color in the Cousins system?
The results of the modeling indicate that the reddened mid-M star can have
a Cousins color that is redder than the late-type star by $\sim0.2$~mag, and 
thus the color transformation does depend on spectral type and reddening. 
Transformations were derived for $A_V=0$ and 3 and the average of the two
was applied to the data set for IC~348. This photometry is used in 
Figure~\ref{fig:ri}. A more precise color correction can be computed
at a particular instrumental color if either the spectral type or the
reddening is known. For the spectroscopic sample discussed in this work,
the spectral types were incorporated into the modeling to produce a second 
set of color corrections. The resulting photometry is provided in Table~1. 
The measurements of $R-I$ of Herbig (1998) agree with the data presented here 
for $R-I<1.6$, but the colors of Herbig (1998) become systematically redder by 
0.3-0.5~mag at $R-I>1.6$. 
On the other hand, as discussed by LRLL, at $R-I>1.6$ the colors
reported by Trullols \& Jordi (1997) become progressively bluer than 
those of Herbig (1998) by 0.3-1.5~mag, and hence bluer than the colors
presented here by up to a magnitude. 
The colors implied by the spectra of the reddest 
sources (e.g., object 405) are more consistent with the photometry reported 
in this work. 

Photometry and coordinates for cluster members with spectral types of M4 or 
later are listed in Table~1. Optical and IR measurements 
of Herbig (1998), Lada \& Lada (1995), and LRLL are included as well. 
The IR photometry of LRLL for sources not found in Table~1 can be obtained
by contacting the author.
Although the IR colors agree for sources in common between the latter studies,
there is an unexplained offset of 0.2~mag in all three bands, where the 
photometry of LRLL is fainter than that of Lada \& Lada (1995). For purposes
of this work, agreement is obtained by arbitrarily subtracting 0.2~mag from the 
measurements of LRLL. Because the completeness limit in the Lada \& Lada (1995) 
survey is near $K=14$, the uncertainties are large ($\pm0.2$~mag)
for the faintest sources in Table~1. The photometry of LRLL 
is deeper ($K\sim16.5$) and should be fairly accurate ($\pm0.05$~mag).

Spectra were obtained for low-mass candidates in IC~348 during a few hours 
of service observations with the Keck~II low-resolution imaging spectrometer 
(LRIS; Oke et al.\ 1995) on 1998 August 7. 
The multi-slit mode of LRIS was used with 
the 150~l~mm$^{-1}$ grating ($\lambda_{\rm blaze}=7500$~\AA) and
GG495 order-blocking filter. The maximum wavelength coverage of
LRIS, 3800 to 11000~\AA, was provided in one grating setting
centered near 7500~\AA. One slit mask covered a field of view of 
$5\arcmin\times7\arcmin$ and used slitlets $8\arcsec$ in length and 
$1\farcs2$ in width. This configuration produced a spectral resolution of 
20~\AA.  Two exposures of 25~min were obtained with a single slit mask.
The spectrophotometric standard star was Feige~11, observed through a 
$1\farcs0$ slit.  After subtracting the bias from the frames and flat-fielding
with internal continuum lamps, the spectra were extracted and calibrated in
wavelength with the Ne and Ar lamp spectra.  The 
data were then corrected for the sensitivity function measured from Feige~11.
These observations provided spectra for the low-mass cluster members 405, 611, 
and 613, in addition to several background stars. 

The remainder of the data were collected at the Kitt Peak Mayall 4~meter
telescope. Object 407 was observed with the 4~meter Cryogenic Camera (CryoCam)
on 1998 December 22, while another $\sim70$ sources were observed with the 
4~meter RC Spectrograph (RCSP) on 1998 December 23 and 26. 
Under good weather conditions, I obtained a spectrum of 407 ($I=19.5$)
with the 300~l~mm$^{-1}$ grism ($\lambda_{\rm blaze}=8010$~\AA), OG550 
order-blocking filter, and $1\farcs7$ slit.
The spectral resolution was 12~\AA\ and the exposure time was 40~min.
Additional measurements were made for the spectrophotometric standard 
Hiltner~102 and the spectral type standards LHS~2065 (M9V) and LHS~2243 (M8V). 
During the observations with RCSP, I used the multi-slit mode to obtain
spectra of 8-10 objects in each of several pointings. 
The slit masks projected a circular field of view that was $5\arcmin$ in 
diameter.  The slitlets were $2\arcsec$ in width and the spectral resolution
was 14~\AA. Spectra were obtained with the 
158~l~mm$^{-1}$ grating ($\lambda_{\rm blaze}=7000$~\AA) and the OG570
order-blocking filter. The data from both RCSP and CryoCam provided
wavelength coverages similar to that of the LRIS spectra.
Because of moderate cirrus during most of the RCSP run, 
I observed with several slit masks designed for brighter stars ($I<16$), 
in addition to two masks that targeted faint, low-mass candidates ($I=17$-19.5).
Exposures times ranged between 30 and 45~min. Hiltner~102 and two young 
late-type sources in Taurus, V410~X-ray~5a and V410~X-ray~6, were observed
through a long slit. For each mask and long slit, 
exposures were taken with quartz continuum and He-Ar-Ne lamps. 
The data reduction procedures were the same as for the LRIS data.

\section{Discussion}

\subsection{Identification of Low-Mass Candidates and Determination of
Cluster Membership}

\begin{figure}
\plotfiddle{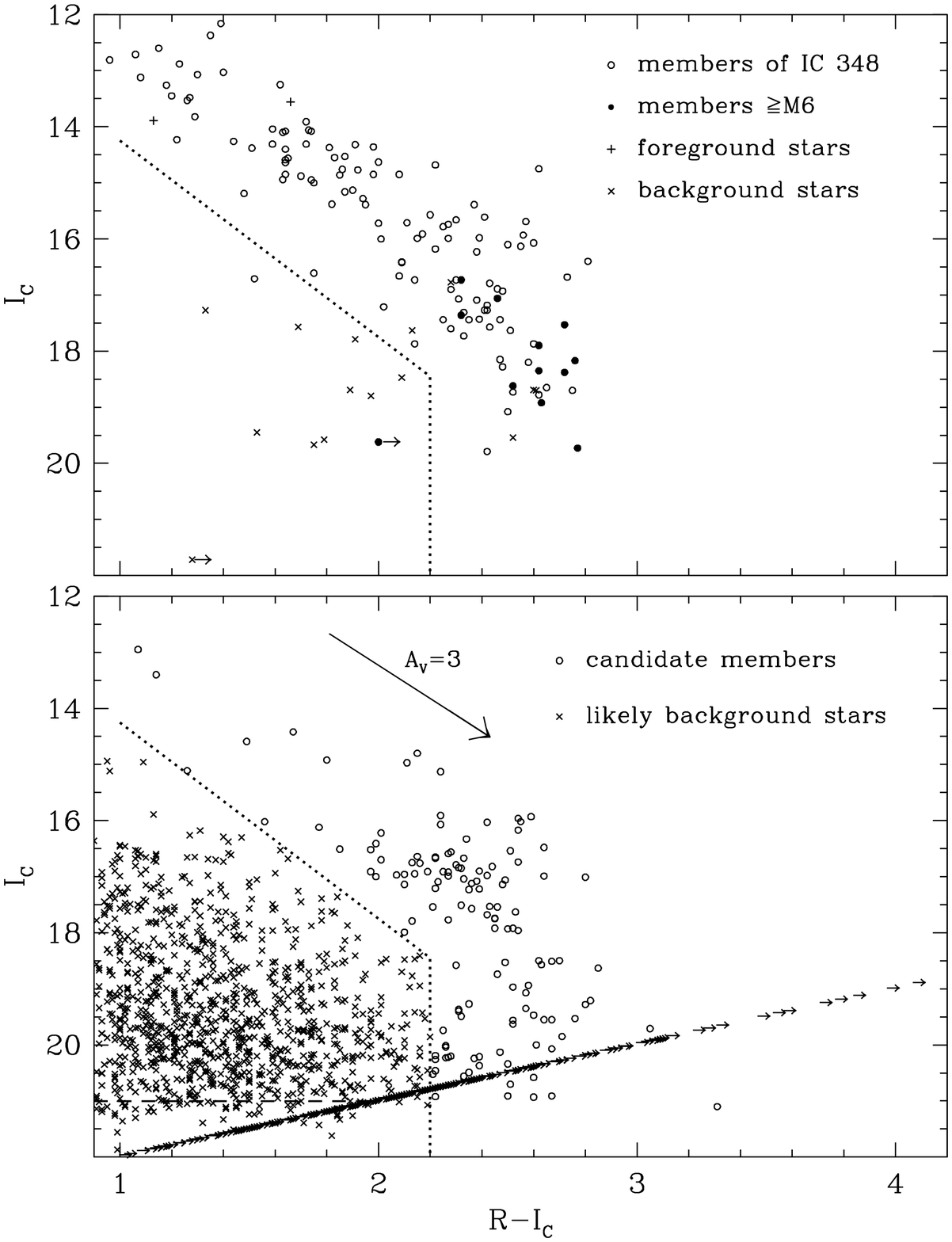}{6.5in}{0}{80}{70}{-245}{-40}
\caption{
$R-I$ vs.\ $I$ for an area of $25\arcmin\times25\arcmin$ towards IC~348. 
The spectroscopic sample is given in the upper panel, where cluster
members, likely substellar members ($\geq$M6), and field stars are indicated. 
The lower panel shows the remaining stars that have not been observed
spectroscopically.
For sources that are saturated in my data ($I\lesssim16.5$), measurements 
of Herbig (1998) are used for the central $7\arcmin\times14\arcmin$ of the 
cluster.  Completeness limits of $I\sim21$ and $R\sim23$ are 
indicated by the dashed line and arrows. 
}
\label{fig:ri}
\end{figure}

A color-magnitude diagram with the new $RI$ photometry for all of IC~348
($25\arcmin\times25\arcmin$) is shown in 
Figure~\ref{fig:ri}. Saturation occurred at $I\sim16.5$ and data of 
Herbig (1998) were used for brighter stars within his $7\arcmin\times14\arcmin$
survey area. At lower masses and cooler temperatures, cluster members should
become redder and fainter. 
The colors of M dwarfs eventually saturate near $R-I=2.4$ for the latest types.
In selecting low-mass candidates to observe through spectroscopy, 
the highest priority was
given to targets near the cluster core to reduce field star contamination and
extend the depth of the previous study of LRLL.
The sources observed spectroscopically by LRLL outlined a locus of likely 
cluster members clearly separated from most of the 
background stars. In addition, IR photometry from LRLL and Lada \& Lada (1995)
was combined with the optical data to help identify low-mass cluster members. 
Since most stars have similar near-IR colors, $J-H$ and $H-K$ provide rough
estimates of extinction, thus distinguishing cool, low-luminosity 
brown dwarfs from reddened, luminous early-type stars (either cluster members 
or background) that appear in the optical color-magnitude diagram.
Unfortunately, at the faintest limits ($I>19$) where the most background
contamination occurs, the available $JHK$ data is sufficiently deep only in the 
cluster core. Actively accreting stars often
have strong ultra-violet and optical excess emission that reduce
the apparent $R-I$ color while leaving $I$ relatively unchanged. Such a 
star could therefore be mistakenly rejected as a background object in the
selection of candidates. 

\begin{figure}
\plotfiddle{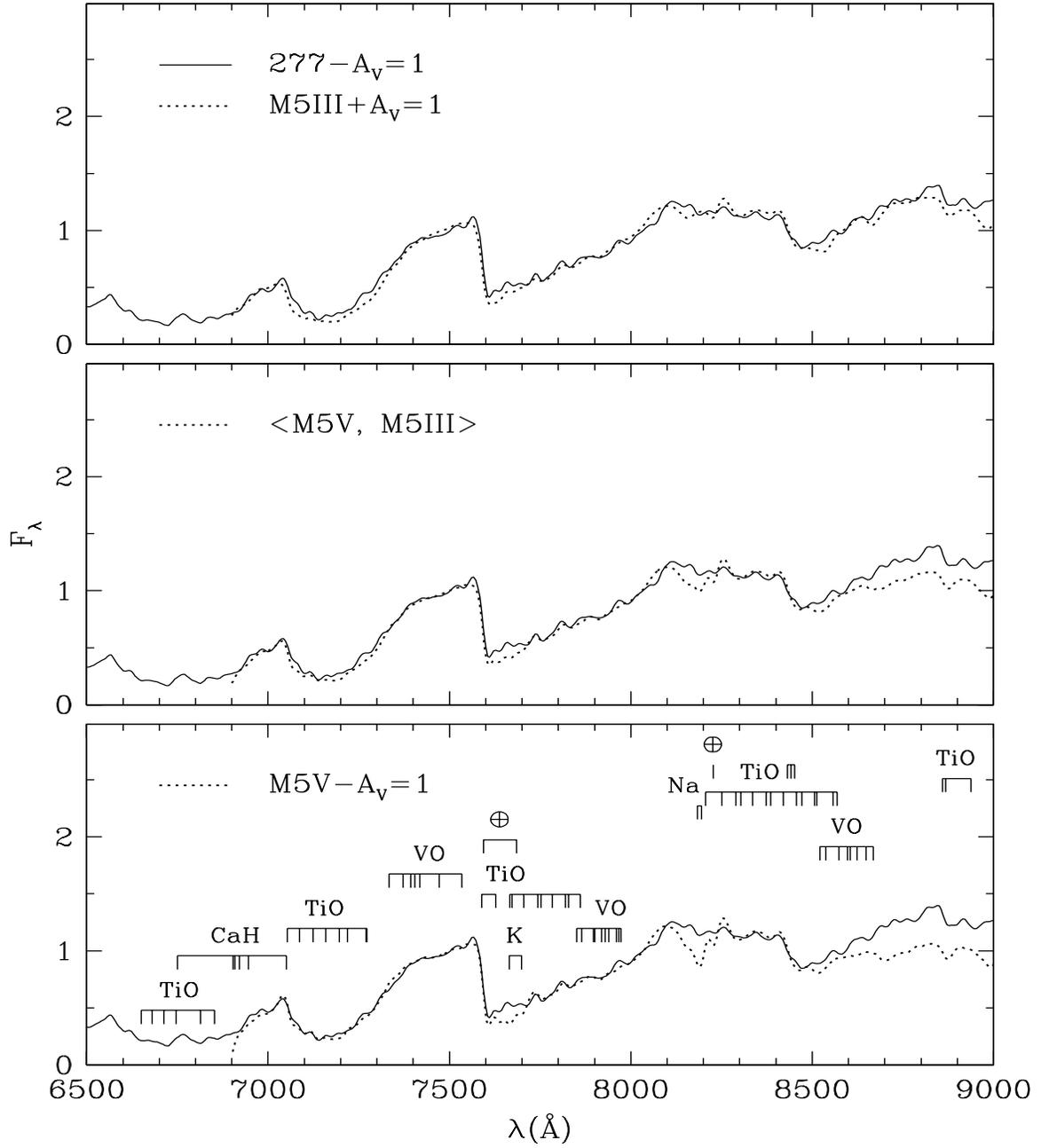}{6.5in}{0}{80}{70}{-245}{-30}
\caption{
The spectrum of object 277, dereddened by $A_V=1$
to approximate the intrinsic spectrum (see \S~\ref{sec:compare}), is shown
with data for M5V, M5III, and an average of the two. For easier comparison 
of the spectral features, the slopes have been matched by adjusting the 
reddenings of the standards.  The young object's intrinsic
spectrum is bluer than M5V, hence a negative extinction must be 
applied to the standard to produce a match.
All data are smoothed to a resolution of 25~\AA\ and normalized at 7500~\AA.
}
\label{fig:m5}
\end{figure}

\begin{figure}
\plotfiddle{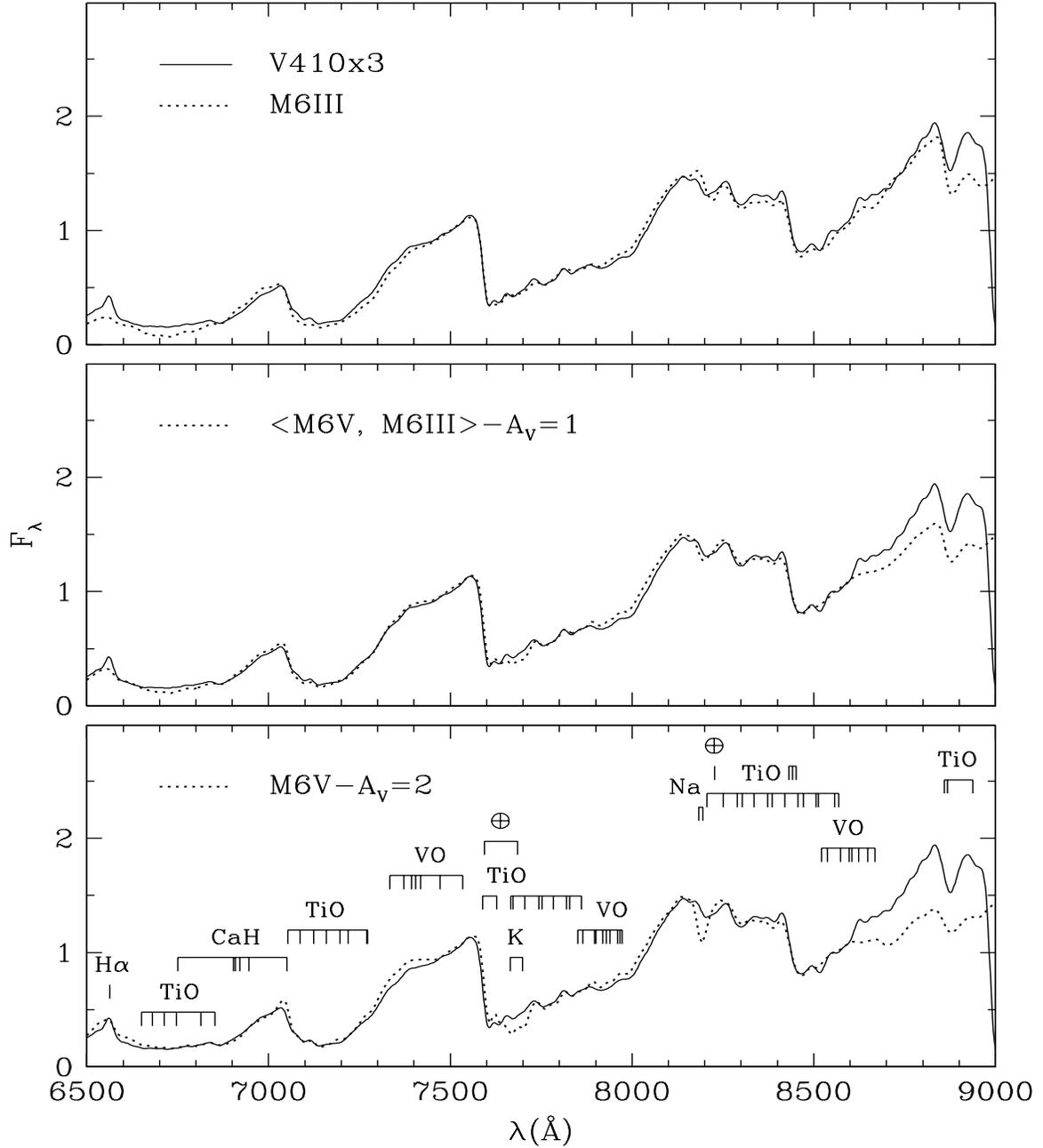}{6.5in}{0}{80}{70}{-245}{-30}
\caption{
The spectrum of V410~X-ray~3 (Luhman et al.\ 1998a), dereddened by $A_V=0.6$
to approximate the intrinsic spectrum (see \S~\ref{sec:compare}), is shown
with data for M6V, M6III, and an average of the two. For easier comparison 
of the spectral features, the slopes have been matched by adjusting the 
reddenings of the standards.  The young object's intrinsic spectrum is bluer 
than M6V and an average of M6V and M6III, hence a negative extinction must be 
applied to the standard to produce a match.
All data are smoothed to a resolution of 25~\AA\ and normalized at 7500~\AA.
}
\label{fig:m6}
\end{figure}

Field stars are easily distinguished from low-mass cluster members with
the spectroscopic data.
Unlike young cool stars, foreground M dwarfs have strong absorption in 
Na and K (see Figs.~\ref{fig:m5}-\ref{fig:m85}), no signs of reddening 
in the spectra or colors, no IR excess emission, and little H$\alpha$ 
emission ($<15$~\AA). Only a few foreground M stars are expected towards
the relatively small area covered by IC~348 (see Herbig 1998). Most 
background stars are rejected by their positions
in the optical color-magnitude diagram. Ones that fall within the locus
of cluster members are background giants, which
exhibit spectra that differ greatly from late M cluster members.

The combined new and published spectroscopic samples are shown in the top panel 
of Figure~\ref{fig:ri}. Reddened low-mass stars belonging to the 
cluster are scattered among the likely brown dwarfs ($\geq$M6).
Because of the compact nature of the cluster, there is very little 
contamination from foreground or background stars within the locus of 
cluster members. As demonstrated in the lower panel of Figure~\ref{fig:ri}, 
there are many likely low-mass cluster members that remain to be observed
spectroscopically.

\subsection{Spectral Types}

\subsubsection{Method of Classification}
\label{sec:method}

For the classification of sources in the new spectroscopic sample,
spectra of standard dwarfs and giants are taken from Kirkpatrick, 
Henry, \& McCarthy (1991), Henry, Kirkpatrick, \& Simons (1994), and
Kirkpatrick, Henry, \& Irwin (1997). The spectral types are 
represented by averages of one or more stars. For M4V-M6V, they 
consist of: M4V=GL~213, GL~275.2A, and GL~402; M4.5V=GL~234AB and GL~268; 
M5V=GL~51 and GL~866AB; M5.5V=GL~65A and G~208-44AB; M6V=GL~406.
The stars that are used for M7V-M9V and M5III-M9III are listed in 
Luhman et al.\ (1998a) and LLR.  There are 
some differences in the band strengths among the standards of a given
spectral type, particularly at M8 and M9 where the VO changes very
rapidly. These fluctuations correspond to $\sim0.25$~subclass. Hence, the
exact spectral type depends slightly on the choice of standards.

As illustrated in 
Figs.~\ref{fig:m5}-\ref{fig:m85}, TiO and VO give rise to several distinctive
absorption features that are the primary indicators of spectral type.
The spectrum of each young object was compared to the dwarfs, giants, and 
averages of the two (each normalized at 7500~\AA)
at various spectral types and reddenings until a 
best match was achieved. After this initial classification of the sample, 
all spectra at the same spectral type were dereddened to match the least 
reddened spectrum. These data were compared closely and minor adjustments 
($\sim0.25$~subclass) were made in the classifications, until a point was 
reached where objects of a given spectral type were identical 
within the noise. Because of the rapid change in optical features with
lower temperatures, the relative classifications are quite precise. 
Typical uncertainties are $\pm0.25$~subclass for objects with one spectral
type listed in Table~1. A larger range of possible spectral types is given
for some sources. Even for data of very low signal-to-noise, the uncertainties
in spectral types were no more than $\pm1$~subclass.
Spectra previously classified in LRLL have been included with the new sample 
when deriving spectral types, some of which have been revised slightly 
from those reported by LRLL. Table~1 lists all known objects in IC~348 with
classifications of M4 or later. New spectral types for earlier sources 
will be presented in a future study.  The IR types from LRLL are given 
in Table~1 as well, in addition to optical types of Herbig (1998) for 
the two sources that were not in my optical sample.  A spectral 
type of M5.5 was measured for the two sources in Taurus, V410~X-ray~5a and
V410~X-ray~6, compared to previous optical classifications of M5 for both 
objects (Strom \& Strom 1994) and M5.5 for 5a (Brice\~{n}o et al.\ 1998).

\subsubsection{Comparison to Dwarfs and Giants}
\label{sec:compare}

While the relative spectral types of the sample in IC~348 are precise, 
the absolute classifications require further attention. 
The late-M spectroscopic standards defined to date consist of field dwarfs and
giants.  The optical spectra of these standards differ between luminosity 
classes for a given spectral type. Hence, classifications
of young M objects can depend on the choice of dwarfs or giants
as the standards and on the wavelength range considered.
In previous studies, I have used spectra of field dwarfs 
and giants and averages of the two in the classification 
of a small number of young late-type objects, finding that the averages 
produce the best match.  The new observations in IC~348 combined with 
the studies of LRLL and LLR provide spectra for a large population of 
young cool sources, including six objects at M8-M8.5. 
Because the young cool sources in this study form well-defined 
spectral classes, I can select objects with low reddening that are
representative of each spectral class. After estimating the extinctions
for these objects, the spectra are then dereddened to their intrinsic form.
These data will be compared to the spectra of 
standard dwarfs and giants and their averages at spectral types of 
M5 through M8.5, thus revealing the spectral behavior of cool PMS objects
and indicating the most suitable choice of standards for their classification.
Similar discussions pertaining to $K$-band spectral features are found in LR98, 
LRLL, and Luhman \& Rieke (1999).

The spectrum of source 277 exhibits the least extinction of the M5 objects
and thus will be used to represent this spectral type upon dereddening. 
The two bluest spectra among slightly earlier and later spectral types are
those of 266 and 163, respectively,
which have similar slopes and are bluer than 277 by $A_V\sim1$. 
It is likely that 266 and 163 have little extinction since they show the 
least reddened spectra out of 26 objects from M4.75-M5.25. An
extinction of $A_V>1$ would imply that their spectra are intrinsically bluer 
than those of both dwarfs and giants and the reddenings derived from $J-H$ 
for 163, 266, and 277 in \S~\ref{sec:colors} are low ($A_V\sim0.5\pm1$).
An extinction of $A_V=1$ is therefore adopted for 277
and used in estimating the intrinsic spectrum. In Figure~\ref{fig:m5}, 
the dereddened spectrum of 277 is shown with the M5 standards. To facilitate
the comparison of individual features between the young star and the standards,
the reddenings of the standards have been adjusted to match their 
7000-8500~\AA\ slopes to the spectrum of 277.
The slope of 277 agrees well with an average of 
M5 dwarf and giant spectra without any such adjustment. Thus, young M5 objects
appear to be redder than giants and bluer than dwarfs. The TiO strength from
7100 to 7300~\AA\ in 277 
is the same as in M5V, while weaker than in M5III, so that the average of the 
two luminosity classes has TiO that is slightly stronger than in 277. 
This young object shows very weak absorption in K and Na, similar
to the giant. Although K and Na are not useful in measuring spectral types, 
they do clearly indicate PMS nature and cluster membership. 
Absorption in VO at 7900 and 8500~\AA\ is not sensitive 
to gravity and is unchanged among the dwarf, giant, and young star.

In a study of the relatively unreddened V410~X-ray~3 ($A_V\sim0.6$), Luhman
et al.\ (1998a) found that the spectral slope from 6500 to 9000~\AA\ of 
this young M6 object was reproduced well by M6III, while bluer than M6V.
In a different comparison in Figure~\ref{fig:m6}, the spectral slopes
of the standard stars and V410~X-ray~3 have been aligned through dereddening, 
as with source 277. The TiO absorption at 7100 to 7300~\AA\ in the giant is 
slightly stronger 
than in the young object. The average of the dwarf and giant matches the CaH,
TiO, and VO between 6900 and 7500~\AA\ quite well, much better than either a
dwarf or giant a alone. The K and Na absorption and overall slope resemble 
a giant more than a dwarf.  The VO at 7900 and 8500~\AA\ in V410~X-ray~3 
is matched by both luminosity classes. 

In the sample for IC~348, there are no spectra showing low reddening and
high signal-to-noise near a spectral type of M7. 
W99 have recently obtained high-quality data for the
relatively unreddened binary components GG~Tau~Ba and Bb. They reported
spectral types of M5$\pm0.5$ and M7$\pm0.5$, respectively, where 
VY~Peg (M7III) was used as the standard in the classifying Bb.
From the optical spectra of GG~Tau~Ba and Bb kindly provided by R. White, 
I have measured spectral types that fall within the uncertainties quoted by 
W99. However, I find that more accurate and precise spectral
types of M5.5$\pm0.25$ and M7.5$\pm0.25$ can be achieved from the data. 
As seen in Figure~3 in W99, the VO absorption at 7900 and 8500~\AA\ is 
stronger in Bb than in either M7V or M7III. A better match to these
features and the remainder of the spectrum is provided by 
an average of M7 and M8 dwarfs and giants.

\begin{figure}
\plotfiddle{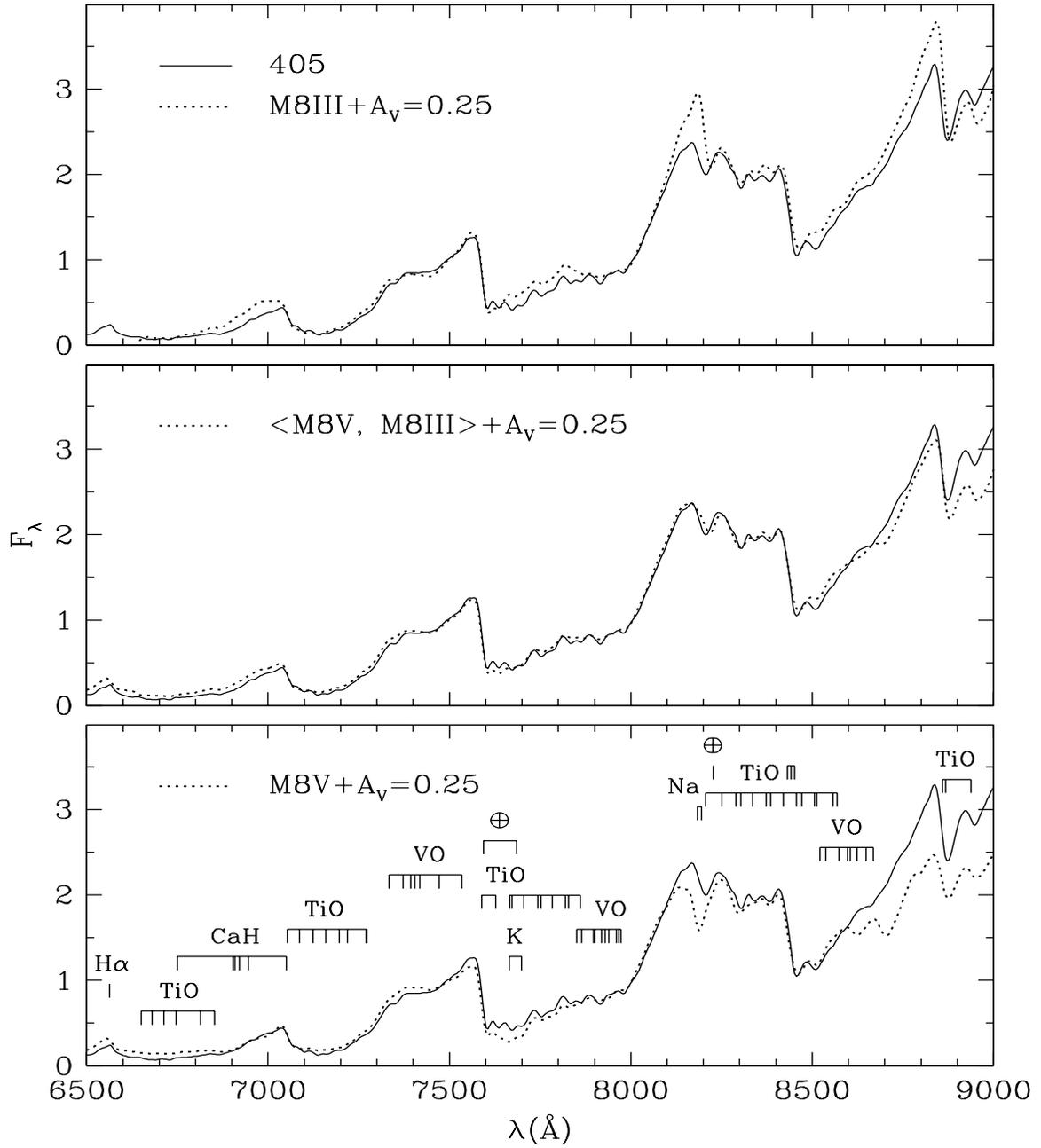}{6.5in}{0}{80}{70}{-245}{0}
\caption{
The standard spectra of M8V, M8III, and an average of the two are 
reddened slightly ($A_V=0.25$) to match the data for source 405. 
All data are smoothed to a resolution of 25~\AA\ and normalized at 7500~\AA.
}
\label{fig:m8}
\end{figure}

\begin{figure}
\plotfiddle{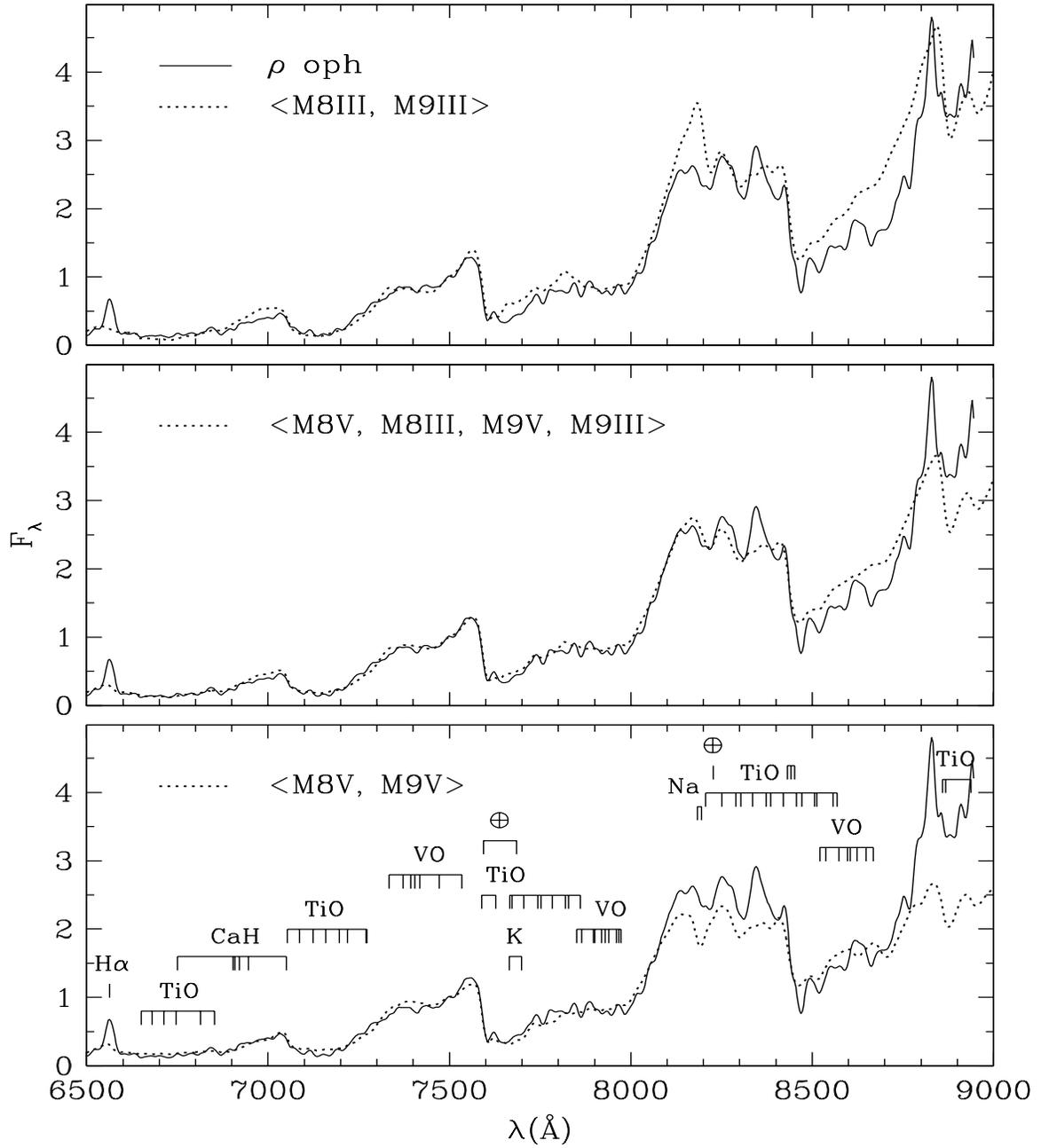}{6.5in}{0}{80}{70}{-245}{0}
\caption{
The spectrum ($A_V<0.7$) of the $\rho$~Oph brown dwarf GY141 is 
shown with data for standard M8 and M9 dwarfs and giants.
All data are smoothed to a resolution of 25~\AA\ and normalized at 7500~\AA.
}
\label{fig:m85}
\end{figure}

The spectrum of source 405 is shown in Figure~\ref{fig:m8} with the M8 dwarf 
and giant standards. By applying a very small amount of reddening ($A_V=0.25$)
to the standard spectra, an optimum match is obtained to the slope of 405 
shortward of 8500~\AA. Late M giants
become progressively redder than dwarfs beyond 8500~\AA\ and 405 is intermediate
between the two.  As demonstrated in Figure~\ref{fig:m8}, the average of
M8V and M8III clearly produces the best agreement with the spectrum of
405, particularly across the structure between 7900 and 8500~\AA.

In a comparison of the spectrum of a young brown dwarf in $\rho$~Oph
(162349.8-242601, GY141) to averages of dwarf and giant spectra at
M7, M8, and M9, LLR found that this object was intermediate between M8
and M9 and assigned a spectral type of M8.5. For a fixed M8.5 spectral class,
the spectrum of this object is now compared to data representing a dwarf,
a giant, and an average of the two in Figure~\ref{fig:m85}. GY141 has strong 
VO at 8500~\AA\ and a red spectral slope that resembles a giant, while the CaH, 
TiO, and VO short-ward of 8500~\AA\ are reproduced well by the dwarf/giant 
average. The spectrum of a young M8.5 object in $\sigma$~Ori is very similar 
to that of GY141 (B\'{e}jar et al.\ 1999), suggesting that GY141 may be 
representative of this PMS spectral type. 

Absorption in VO at 7900 and 8500~\AA\ is a very good indicator of spectral
type in young late-type objects since it is strong, easy to detect in 
faint sources, and does not depend on surface gravity. Most of the other
spectral features have strengths intermediate between dwarf and giant values.
Indeed, straight averages of field dwarf and giant
spectra reproduce the data for cool PMS sources remarkably well. Except for 
differences in reddening, the examples described here are generally 
representative of young ($\lesssim10$~Myr) cool objects in IC~348 and in other 
regions.  Thus, I advocate the use of averages of dwarf and giant standards 
for classification of young objects at M5 and later, with the caveats for 
individual features discussed in this section.
This choice of standards is also practical since the majority of young 
late-type objects discovered to date have been classified in this manner 
(this work, LRLL, LLR). Young sources with little reddening ($A_V<1$) 
and precise classifications should be used as supplemental standards.

\subsection{Colors and Extinctions}
\label{sec:colors}

The typical behavior of the optical and IR colors of young late-type sources 
is investigated in the data for IC~348, followed by a determination of
the best method of deriving extinctions for such a population.
Standard dwarf colors are taken from the compilation of Kenyon \& Hartmann
(1995) for types earlier than M0 and from the young disk populations described 
by Leggett (1992) for types of M0 and later. The IR colors are placed
on the CIT system with relations between the Johnson-Glass and CIT systems
found in Bessell \& Brett (1988). Reddenings are calculated with the
extinction law of Rieke \& Lebofsky (1985).

\begin{figure}
\plotfiddle{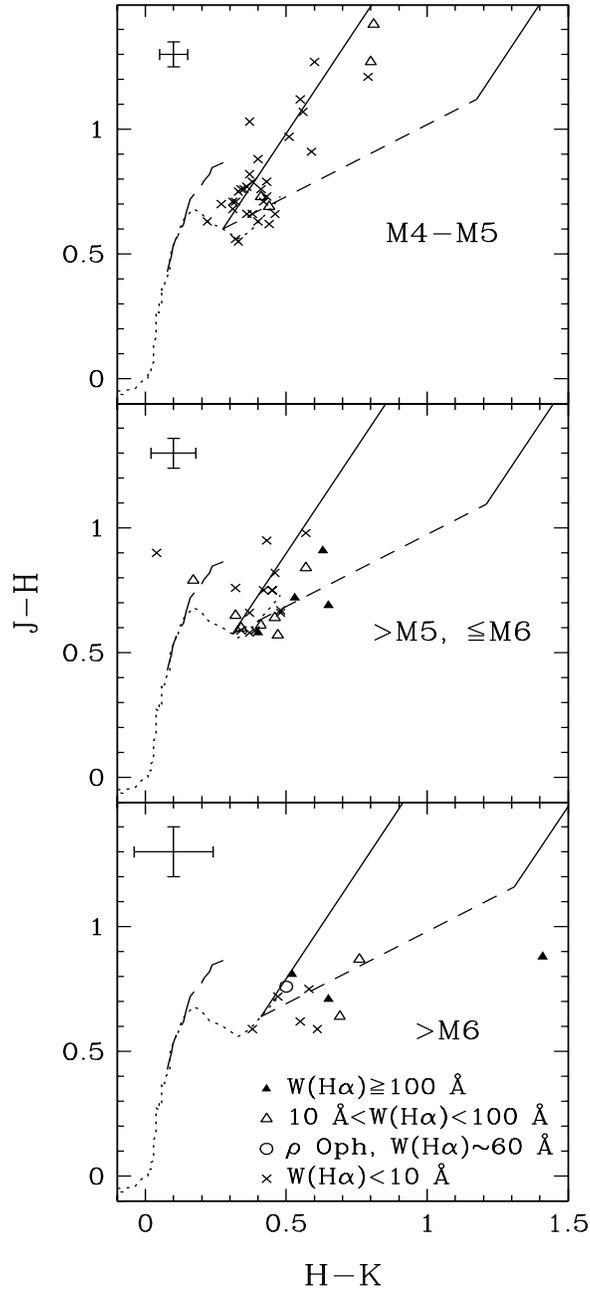}{6.5in}{0}{80}{70}{-245}{-40}
\caption{
$H-K$ vs.\ $J-H$ for late-type members of IC~348 and the $\rho$~Oph brown 
dwarf GY141 sorted by spectral type and H$\alpha$ strength and shown with
typical colors of field dwarfs (dotted line; $\leq$M9) and giants (long 
dashed line; $\leq$M5). 
The locus of classical T~Tauri stars ($\sim$M0) measured by Meyer, Hillenbrand, 
\& Calvet (1997) (short dashed line) and reddening vectors (solid lines)
are plotted with the origin placed at the dwarf colors of each spectral type.
The error bars represent the typical uncertainties in the colors from Lada
\& Lada (1995). More accurate colors from LRLL have been used when available.
}
\label{fig:jhhk}
\end{figure}

The low-mass sources in IC~348 and standard dwarfs and giants are shown in
a diagram of $H-K$ versus $J-H$ in Figure~\ref{fig:jhhk}.
With later M types, $J-H$ colors for dwarfs remain near 0.6-0.7 while 
$H-K$ colors increase from 0.15 to 0.5. Giants depart from dwarf
colors near $J-H=0.6$ and $H-K=0.1$ and approach colors of 1 and 0.3 at the 
latest types (Bessell \& Brett 1988).
For each of the three ranges of spectral types in Figure~\ref{fig:jhhk}, 
a reddening vector originates at the corresponding dwarf colors.
The young systems in IC~348 move progressively to redder $H-K$ with
later types, closely following the dwarf behavior.  The $J-H$ colors in
the least reddened objects are similar to dwarfs and bluer than giants,
indicating that both IR colors of the central stars are dwarf-like.

As with higher mass CTTS, some of the low-mass 
objects in IC~348 show emission in excess of reddened dwarf colors.
Meyer, Calvet, \& Hillenbrand (1997) measured a locus of dereddened IR 
colors for classical T~Tauri stars near M0 and modeled it in terms of
star-disk systems, where the origin of this locus coincides with the unreddened 
colors of an M0 dwarf.  At later M types, they predict that the origin 
will continue to follow the dwarf colors while maintaining a fairly constant
slope. To test this suggestion, a fit to the CTTS locus of Meyer et al.\ (1997)
is given in Figure~\ref{fig:jhhk} with the origin adjusted for each of the 
three spectral types.  As discussed shortly, it may not be possible to measure 
accurate reddenings from optical colors of these young cool objects. Hence, it 
is difficult to deredden the IR colors for comparison to the CTTS locus. 
The large photometric uncertainties and the small number of sources
with significant $H-K$ excesses provide further obstacles in determining
the intrinsic colors of low-mass CTTS. However, it is apparent that 
at the latest types the
average observed colors fall below the CTTS locus, a trend that would become
more pronounced if the colors were dereddened. In other words, these systems 
seem to have significant color excesses at $H-K$ but not $J-H$.
Relative to CTTS at higher masses, this behavior would suggest cooler
emitting regions with respect to the central objects, possibly due to 
cooler disks, larger inner holes, or contributions from material in 
infalling envelopes. The latter emission source is particularly likely 
in object 407, which has a higher $H-K$ excess than can be easily explained
by star-disk systems (Meyer, Calvet, \& Hillenbrand 1997). As seen in
Figure~\ref{fig:jhhk}, the systems with higher H$\alpha$ emission tend
to show larger excesses at $H-K$, which is expected for more actively
accreting disks. It also appears that on average the
latest types show the largest excesses. However, the uncertainties in
the colors are largest for the faintest, coolest objects. For instance,
the anomalous colors of 432 ($J-H=0.9$, $H-K=0.04$) reported by Lada \& Lada 
(1995) are likely due to the uncertain $K$-band measurement ($\pm0.39$~mag).
A more conclusive and detailed analysis of these various issues requires 
more accurate $JHK$ photometry and data at longer wavelengths. 

As demonstrated by the $J-H$ color excesses in Figure~\ref{fig:jhhk} 
($A_J=2.63 E(J-H)$), the sources at M4-M5 exhibit $A_J=0$-2 while the later 
types have $A_J\lesssim1$, which is a reflection of the completeness limit of 
the spectroscopy and selection against reddened late-type objects.
The M4-M5 stars with $A_J<1$ are saturated in the new $RI$ photometry.
To examine $R-I$ and $I-J$ for the remaining sources, the colors have
been dereddened with the extinctions derived from $J-H$ assuming intrinsic
colors of dwarfs.  The dereddened $R-I$ colors are bluer than dwarf values 
by 0.2-0.6~mag for objects at M4-M5, which is a tendency towards giant-like
colors. At later types, the young sources
have colors similar to dwarfs and giants, which have comparable 
Cousins $R-I$ colors for M6 and later. (It is redward of Cousins $I$ 
where late M giants become redder than dwarfs.)
This behavior is consistent with the
spectra, where the slopes across the Cousins $R$ and $I$ bands are bluer 
than in dwarfs for M4-M6 and similar to both dwarfs and giants at later types.
Because the cooler objects have low extinctions
and the non-saturated M4-M5 stars are reddened, the results of this
comparison are susceptible to systematic effects from the color correction.
However, careful modeling of the color transformation as 
a function of spectral type and reddening (\S~\ref{sec:obs}) indicates that
the optical colors are fairly accurate ($\pm0.1$) and the blue dereddened 
colors at M4-M5 should be a real effect. 
Comparing the optical and IR data, the dereddened $I-J$ colors are
redder than dwarf values by 0, 0.3, and 0.6~mag at M4-M5, $>$M5-M6, and $>$M6.
As discussed in \S~\ref{sec:obs}, the IR photometry of LRLL was 
offset by $-0.2$~mag to agree with the data of Lada \& Lada (1995). If the 
reverse adjustment is made, then these color differences in $I-J$ are
reduced by 0.2~mag. A noticeable departure from dwarf colors would remain
at the latest types, with $I-J$ intermediate between that of dwarfs
and giants. The spectra of the latest objects rise rapidly beyond the
$I$-band in a similar fashion as giants, confirming the behavior of $I-J$.
Color anomalies from $B$ through $J$ have also been observed in earlier type
(K7-M1) PMS stars by Gullbring et al.\ (1998). Because dwarf and giant 
colors are similar in this spectral type regime, they suggest that cool
companions and star spots are instead the probable causes.

To derive extinctions for stars earlier than mid-M, LRLL used $E(R-I)$ and
assumed intrinsic colors of dwarfs, with the constraint that the extinction
could not produce a dereddened $J-H$ much bluer than in dwarfs. When $R$ and
$I$ were not available, the stars were dereddened to the dwarf colors of $J-H$.
For the three latest sources in LRLL, the extinctions were estimated from 
$I-K$ by assuming dwarf colors since this was the most accurate color for
these faint objects. The new results in this study indicate that the intrinsic 
near-IR colors of young late-type objects are probably dwarf-like while $I$ 
relative to $J$, $H$, and $K$ appears to be redder than in dwarfs. The $R-I$ 
colors of young
objects also seem to deviate from those of dwarfs, making it difficult to 
use $R$ or $I$ in measuring extinctions. Possible systematic effects in the
optical color transformation for reddened late-type stars is an added concern. 
Unlike optical data, near-IR photometry is not susceptible to significant 
color corrections and the colors are consistently dwarf-like, thus extinctions 
are derived in this work by dereddening $J-H$ to dwarf colors.  Unfortunately,
much of the near-IR data for IC~348 suffers from large photometric errors at the
faintest objects. The more accurate IR data from LRLL is used when possible
and the resulting extinctions are given in Table~1, which are generally
consistent with the reddenings implied 
by the spectra.  Since most sources exhibit $H-K$ excesses of $<0.2$~mag, the 
contamination of the $J-H$ color by excess emission should be negligible. 
Object 407 has a large $H-K$ excess, thus an extinction cannot be 
confidently derived with $J-H$. Cousins $R-I$ colors are similar between
dwarfs and giants at the latest M types, hence this color is used in 
estimating a reddening of $A_J\sim1$, which is consistent with the appearance
of the spectrum.
Since LRLL could not measure $H$-band data for source 413, the $J-K$ color
was used to measure the extinction.

\subsection{Bolometric Luminosities and Effective Temperatures}
\label{sec:teff}

To minimize contamination from UV and IR excess emission,
the $I$ and $J$ bands are generally preferred for estimating the bolometric
luminosities of young stars. Both the photometry and the spectra indicate that 
$I-J$ may be redder in young cool objects than in dwarfs. Because 
$J-H$ and $J-K$ are dwarf-like, $J$ is likely a better
choice than $I$ for measuring luminosities. The dwarf-like
near-IR colors and the fact that most of the luminosity is released in
$J$ through $K$ suggest that the bolometric corrections for dwarfs may be
satisfactory approximations for these young sources. Hence, the 
luminosities reported in Table~1 have been calculated by applying the 
dwarf bolometric corrections to the dereddened $J$-band data and assuming
a distance modulus of 7.5.
Bolometric corrections are taken from Kenyon \& Hartmann (1995) for 
$<$M6 and compiled from Bessell (1991), Monet et al.\ (1992), Tinney, 
Mould, \& Reid (1993), and Leggett et al.\ (1996) for $\geq$M6. 
Given the uncertainties in the photometry, reddenings, bolometric
corrections, and distance, the typical errors in the bolometric luminosities
are $\pm0.08$ to 0.13 in log~$L_{\rm bol}$ from early K to the latest M types.

Theoretical mass tracks at low masses are mostly vertical in the H-R diagram, 
thus the conversion of spectral types to effective temperatures directly
influences the mass estimates for young late-type objects. 
LR98 discussed the differences between the available temperature 
scales for M dwarfs, finding that the comparison of synthetic spectra to
observed data by Leggett et al.\ (1996) produced a scale in good agreement 
with measurements for the eclipsing spectroscopic binaries YY~Gem and CM~Dra.
Because the latest type in the study of Leggett et al.\ (1996) was M6.5,
LR98 extrapolated the scale to later types assuming the same offset between the
M6-M9 subclasses as found in a previous generation of spectral modeling 
by Kirkpatrick et al.\ (1993).  This version of the scale of Leggett et 
al.\ (1996), given in Table~2, is consistent with newer modeling of colors 
at the latest M types by Leggett, Allard, \& Hauschildt (1998). 
Since PMS stars are intermediate in surface gravity between dwarfs and 
giants, it is useful to consider the temperature scale for giants as well. 
The giant temperature scale in Table~2 is from the fit provided by van~Belle 
et al.\ (1999) for types earlier than M7 and from Perrin et al.\ (1998) 
and Richichi et al.\ (1998) for M7 to M9. The temperatures of giants are warmer
than those of dwarfs by 200-400~K, as illustrated in Figure~\ref{fig:scale}.

In previous studies of young populations composed primarily of objects at mid-M 
and earlier, LR98, LRLL, Luhman \& Rieke (1999) applied the dwarf temperature
scale, while Luhman et al.\ (1998a) and LLR explored the effect of a 
warmer, more giant-like scale on mass estimates for two late M objects. 
For the late-type sources in IC~348 in Table~1, spectral
types are converted to temperatures with the dwarf scale. In \S~\ref{sec:hr},
it will be determined if this scale or one intermediate between those of
dwarfs and giants can produce agreement between the available theoretical 
isochrones and the empirical isochrones formed by the locus of stars in 
IC~348 and the components of GG~Tau. 

\subsection{H-R Diagram}
\label{sec:hr}

Using the temperatures and luminosities for objects in Table~1 ($\geq$M4) and 
in the study of LRLL ($<$M4), an H-R diagram is generated for 
the low-mass population in IC~348 and shown with three sets of theoretical 
evolutionary tracks in Figs.~\ref{fig:hr1} and \ref{fig:hr2}.
The models include those by D'Antona \& Mazzitelli (1997) (hereafter DM97),
Burrows et al.\ (1997), and Baraffe et al.\ (1998) (hereafter B98) 
with a mixing length of $\alpha=1.9$.  
The models of Burrows et al.\ (1997) 
shown here are not the same as the older 1997 suite of 
Burrows presented in LLR, LRLL, and Luhman et al.\ (1998a).
Burrows et al.\ (1997) calculated their own synthetic atmospheres,
whereas the NextGen atmospheres of Allard were used in the previous generation
of models.  With evolutionary models, the data for 
a young population can be interpreted in terms of masses and ages and
combined into a mass function and a star formation history. 
However, even the most recent models imply significantly different masses
and ages for the same objects. In addition, the temperature scale for young
cool stars may differ from the dwarf conversion.  Therefore, meaningful 
estimates of masses and ages require observational constraints of the 
models and the temperature scale.

Luhman (1998) briefly reviewed some of the observational tests of the models 
at ages of $\geq0.1$~Gyr, which included the independently
measured masses, radii, and temperatures of CM~Dra and YY~Gem and the 
empirical isochrones observed in the Pleiades and in globular clusters.
In a more detailed study, Stauffer, Hartmann, \& Barrado (1995) compared
each set of model isochrones to the Pleiades locus and determined whether the 
temperature scale could be adjusted to produce agreement. 
While the models at evolved ages are similar to each other and in reasonable 
agreement with observations, at ages of $<10$~Myr they differ greatly 
and lack definitive constraints (e.g., eclipsing binaries). 
As a crude test at young ages ($\sim1$~Myr), Luhman (1998) calculated the 
IMF of L1495E with each set of tracks to search for anomalously large 
deviations from the field mass function. 

\subsubsection{Coevality as a Test of the Models}

Coevality can be utilized as a constraint of the models at the youngest ages
in the same way that the Pleiades locus acts as a test near 100~Myr.
Binary systems should be one source of coeval stars, as explored by
Hartigan, Strom, \& Strom (1994) and Prato (1998).
More recently, W99 have measured spectral types and photometry
for the members of the quadruple system GG~Tau. An H-R diagram was
constructed from this data, where the presumably coeval components should 
form an empirical isochrone extending across a large range of spectral types
(K7-M7). W99 examined several sets of 
models and found that the calculations of B98 with
a mixing length of $\alpha=1.9$ could be combined with a temperature scale
intermediate between those of dwarfs and giants to produce coeval ages 
for the components of GG~Tau. These models also implied masses of DM~Tau
and GG~Tau~Aa+Ab that were in rough agreement with dynamical estimates.
Using revised spectral types for GG~Tau~Ba and Bb (\S~\ref{sec:compare})
and the reddenings and $J$-band photometry measured by W99,
the luminosities of GG~Tau have been calculated in the same manner as for the
sources in IC~348. The luminosities agree within the uncertainties with those
of W99. The revised measurements for GG~Tau and the new data
for IC~348 are combined in the following analysis.

\subsubsection{GG~Tau and IC~348 as Empirical Isochrones}

A test of the model isochrones with data for GG~Tau and IC~348 is based
on the assumptions that the components of GG~Tau are coeval and that 
luminosities are precise indicators of age. Considering the stellar densities
in Taurus and separations within GG~Tau, the members of this multiple system
have probably formed through fragmentation rather than capture or disk 
instabilities, thus coevality is likely (Ghez, White, \& Simon 1997). 
When the populations of young clusters
are placed on H-R diagrams, the distribution of luminosities at a given
temperature is generally interpreted in terms of an age spread. However, 
other factors in addition to age, such as binarity and star spots, may 
significantly influence the observed luminosities.  If true, the measured
luminosity of a particular star would only indicate a crude age, while 
the average distribution of stars in a cluster should still reflect the
age of the cluster and act as an empirical isochrone. Thus, a comparison
of GG~Tau to the locus of stars in IC~348 can test both the coevality of
the multiple system and the precision with which luminosities trace age. 
As illustrated in Figs.~\ref{fig:hr1} and \ref{fig:hr2}, the components
of GG~Tau form a line parallel to the population 
in IC~348. If the observed luminosities of young stars were dominated 
by factors other than age, the components of GG~Tau should have luminosities
randomly drawn from the range of values found in young clusters, and this 
is clearly not the case.  Hence, it is indeed likely that the luminosities do  
reflect the age of GG~Tau and that the components are coeval.

For each set of models, I will examine if the inferred age for GG~Tau and
average age and age spread for IC~348 are constant as a function of mass, 
whether a temperature scale intermediate between those of dwarfs 
and giants can improve agreement, and review other observational constraints 
discussed by LR98, Luhman (1998), and W99.
The sample in IC~348 is representative down to M5-M6 and biased towards 
less reddened and more luminous objects at later types. Consequently,
for stars earlier than M6, the average and upper and lower boundaries of the
locus can be compared directly to the model isochrones. Later than M6, the
upper envelope should be representative, and the average locus and lower
envelope can be treated as upper limits. 

\subsubsection{D'Antona \& Mazzitelli 1997}

The models of DM97 are fairly consistent with the
stellar parameters derived for the eclipsing
binaries CM~Dra (0.22~$M_{\odot}$) and YY~Gem (0.6~$M_{\odot}$). 
The predicted radii agree with the values measured for CM~Dra while different
by two sigma from those of YY~Gem (see Luhman 1998). The temperatures and 
luminosities are within two sigma and one sigma of the measurements 
for CM~Dra and YY~Gem, respectively.  
As seen in Figure~\ref{fig:hr1}, when the Pleiades brown dwarfs
Teide~1 and Calar~3 are placed on the H-R diagram
with the dwarf temperature scale, the age inferred from DM97
agrees with the age of the Pleiades (125~Myr; Stauffer, Schultz
\& Kirkpatrick 1998). The models are also consistent with the masses (65~$M_J$;
Basri \& Mart{\'\i}n 1998) and ages of the binary components of PPL~15.
An additional test of the models is provided by the dynamical mass measured
for the system of GG~Tau~Aa and Ab through observations of the 
circumbinary disk ($1.28\pm0.08$~$M_{\odot}$; Dutrey,
Guilloteau, \& Simon 1994; Guilloteau, Dutrey, \& Simon 1999). 
Similar estimates are available for GM~Aur ($0.84\pm0.05$~$M_{\odot}$; Dutrey 
et al.\ 1998) and DM~Tau ($0.47\pm0.06$~$M_{\odot}$; Guilloteau \& Dutrey 
1998). For DM97, W99 found that the inferred masses
of GG~Tau~A (0.80~$M_{\odot}$) and GM~Aur 
(0.51~$M_{\odot}$) are lower than the observed values, while the 
predicted mass of DM~Tau (0.44~$M_{\odot}$) is consistent with the data.
However, because of the uncertainty in the inclination of GM~Aur, Dutrey et 
al.\ (1998) cannot rule out a mass of 0.6~$M_{\odot}$, which is close to the
mass implied by DM97. 

The H-R diagram in Figure~\ref{fig:hr1} indicates
that the average age and age spread for IC~348 with the models of DM97 
is fairly constant as a function of mass, except at 0.05-0.15~$M_{\odot}$
where the isochrones rise relative to the observed locus. The same
trend is seen for GG~Tau. The components Aa, Ab, and Bb are coeval on the model
isochrones while Ba is older. The only change in the temperature
scale that could produce agreement between the empirical and theoretical
isochrones is adjusting spectral types of M4-M6 to cooler temperatures 
without changing the scale at other types. Such a conversion would be 
discontinuous and cooler than the scales of dwarfs and giants and 
therefore is not a reasonable option. The models of DM97 are most compatible
with the dwarf temperature scale adopted in this work, supporting the accuracy
of the IMFs derived in previous studies of LRLL and Luhman \& Rieke (1999)
that used this combination of tracks and temperature scale. However, 
DM97 may underestimate masses above 0.5~$M_{\odot}$,
possibly introducing a systematic error in the IMFs at intermediate masses.

\subsubsection{Burrows et al.\ 1997}

Since the models of Burrows et al.\ (1997) include only masses of 
$\leq0.1$~$M_{\odot}$, they cannot be tested directly against the 
data for CM~Dra and YY~Gem or the dynamical mass estimates of GG~Tau A, 
DM~Tau, and GM~Aur.  Model~X of Burrows et al.\ (1993) reproduces 
the radii of the CM~Dra components but predicts temperatures and 
luminosities that fall outside of the uncertainties of the measurements 
by two and three sigma, respectively
(Luhman 1998).  At 0.1~$M_{\odot}$, the radius, temperature, and
luminosity at the main sequence calculated by Burrows et al.\ (1997) are 
very similar to the values predicted in Model~X, thus the newer models probably 
do not agree with the data for CM~Dra either.  As with DM97, the calculations 
of Burrows et al.\ (1997) are consistent with the data for the Pleiades brown 
dwarfs.  Given the limited range of masses calculated by Burrows et al.\ (1997),
a comparison of the theoretical isochrones to the locus in IC~348 and GG~Tau is 
not conclusive. The models imply that most of the objects in IC~348 have
ages less than 1~Myr, much younger than expected from other evidence, such
as the fraction of sources exhibiting IR excess emission (LRLL; Lada \& Lada
1995). There is also
a tendency towards younger ages at higher masses in Figure~\ref{fig:hr1},
and no temperature scale between dwarfs and giants removes this implied age
gradient.  

\subsubsection{Baraffe et al.\ 1998}

The radius, temperature, and luminosity as a function of mass predicted 
by B98 for a main sequence star closely matches the 
calculations of DM97 above 0.2~$M_{\odot}$. Thus, the comparison of DM97 to 
the data for CM~Dra and YY~Gem applies to B98 as well.
In the H-R diagram in Figure~\ref{fig:hr2}, the 100~Myr isochrone of B98
is within the uncertainties of data for the Pleiades brown dwarfs 
Teide~1 and Calar~3. However, if PPL~15 is an equal mass binary, 
then the system luminosity shown in the H-R diagram should be reduced by 
0.3~dex to reflect the individual components. At this new position PPL~15,
falls below the 
100~Myr isochrone and above the hydrogen burning limit, which is not
consistent with the age of the Pleiades or the substellar mass of 65-80~$M_J$
implied by the binary data (Basri \& Mart{\'\i}n 1998) and 
Li measurements (Basri, Marcy, \& Graham 1996).

The theoretical PMS evolution of stars at intermediate masses 
($>0.6$~$M_{\odot}$) is sensitive to the treatment of convection 
(Chabrier \& Baraffe 1997). Before the study of B98 was published, 
models with a mixing length parameter of $\alpha=1.0$ were made available
for use in the work of LR98, LLR, Luhman (1998),
and Luhman et al.\ (1998a). B98 subsequently concluded that calculations with 
$\alpha=1.9$ reproduced the properties of the Sun. 
Compared to the observed masses of $1.28\pm0.08$~$M_{\odot}$, 
0.6-0.9~$M_{\odot}$, and $0.47\pm0.06$~$M_{\odot}$ for GG~Tau~A, GM~Aur, and
DM~Tau, W99 derived B98 masses of 2.00, 1.06, and 0.67~$M_{\odot}$ 
for $\alpha=1.0$ and 1.46, 0.78, and 0.64~$M_{\odot}$ for $\alpha=1.9$.
Not only do the models with $\alpha=1.0$ fail to work for the Sun, but they
also significantly overestimate masses in young 
solar-mass stars, which was indicated previously by LR98 and Luhman (1998)
in examining the IMF of L1495E. On the other hand, the calculations with 
$\alpha=1.9$ produce fairly good agreement with the data for
solar-mass stars at both the main sequence and the earliest stages.

\begin{figure}
\plotfiddle{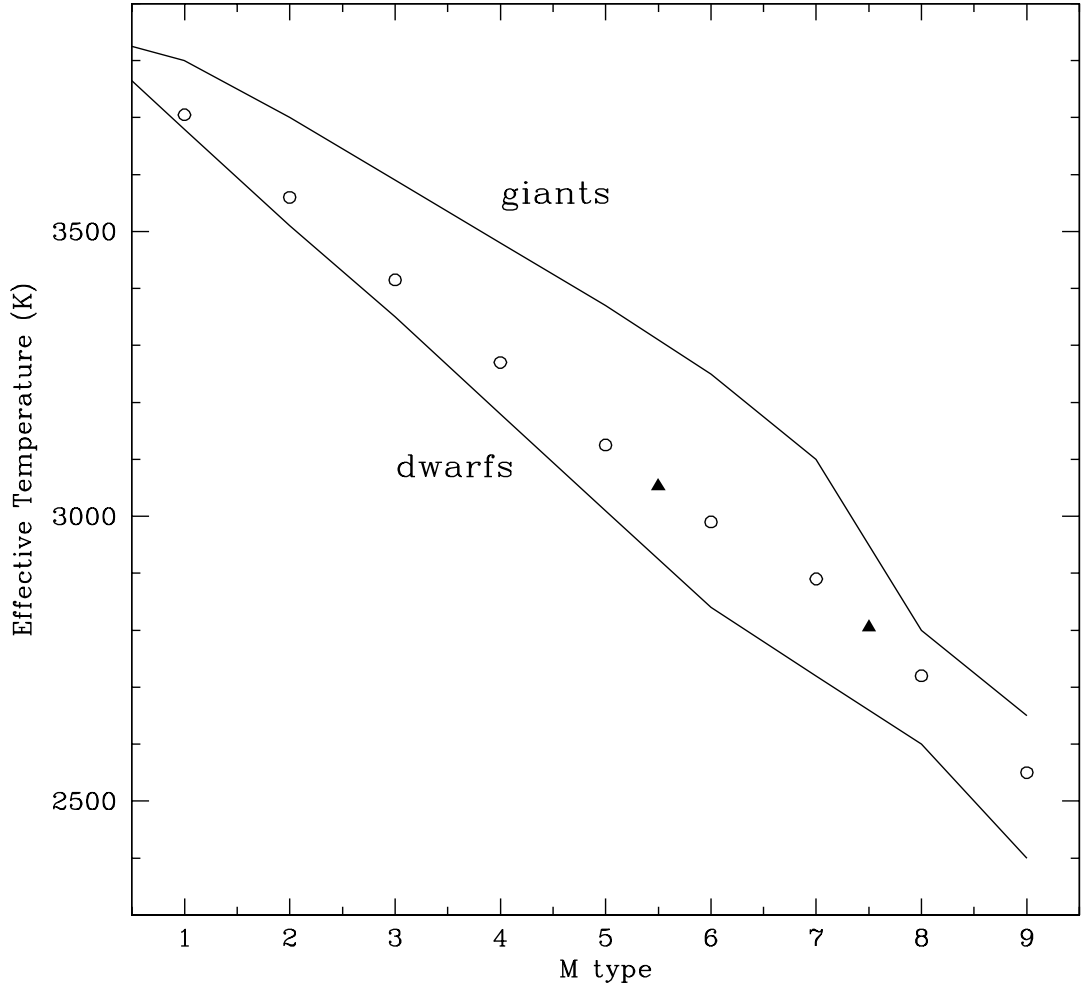}{6.5in}{0}{80}{77}{-245}{-60}
\caption{
The temperature scales for cool dwarfs (Leggett et al.\ 1996;
see text) and giants (Perrin et al.\ 1998; Richichi et al.\ 1998; van~Belle 
et al.\ 1999) listed in Table~2. If the four components of GG~Tau are coeval, 
then the models of B98 imply temperatures 
for GG~Tau~Ba and Bb (solid triangles) that are between dwarf and giant values. 
An intermediate temperature scale that is consistent with these 
results has been constructed (open circles) and is used in the H-R diagram
in the lower panel of Figure~\ref{fig:hr2}. 
}
\label{fig:scale}
\end{figure}

\begin{figure}
\plotfiddle{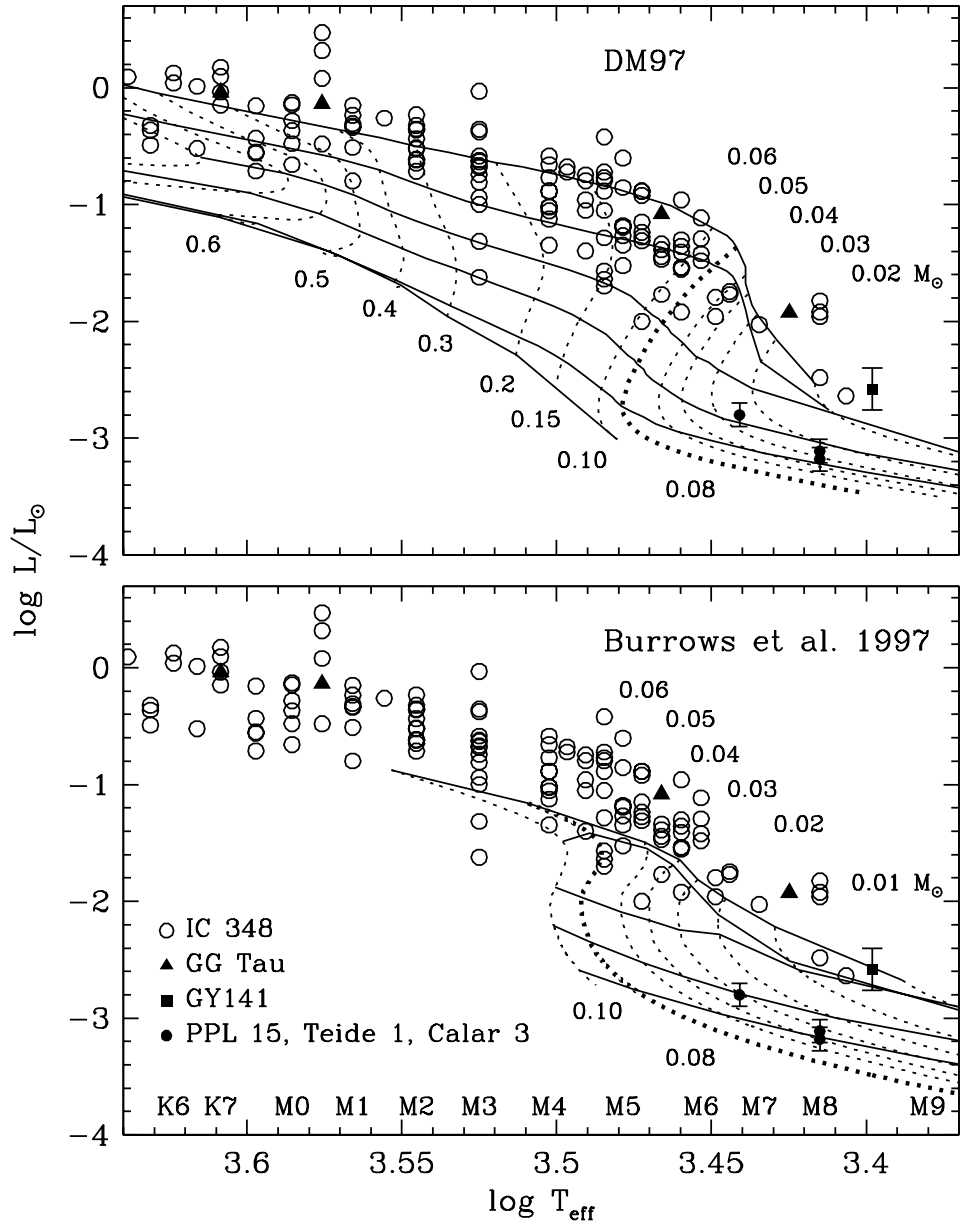}{6.5in}{0}{100}{100}{-306}{-30}
\caption{
H-R diagram for all known late-type sources in IC~348 assuming a 
distance modulus of 7.5 with the evolutionary models of DM97 and 
Burrows et al.\ (1997), where the horizontal solid lines are isochrones 
representing ages of 1, 3, 10, 30, and 100~Myr and the main sequence, from 
top to bottom.  The latter calculations differ from those of Burrows shown 
in LLR and LRLL (see text). 
The four components of the multiple system GG~Tau (White et al.\ 1999) are
expected to be coeval. The $\rho$~Oph brown dwarf GY141 and the
Pleiades brown dwarfs PPL~15, Teide~1, and Calar~3 are also given for reference.
Spectral types have been converted to effective temperatures with a dwarf
temperature scale. 
Uncertainties in spectral types are typically $\pm0.25$~subclass. 
}
\label{fig:hr1}
\end{figure}

\begin{figure}
\plotfiddle{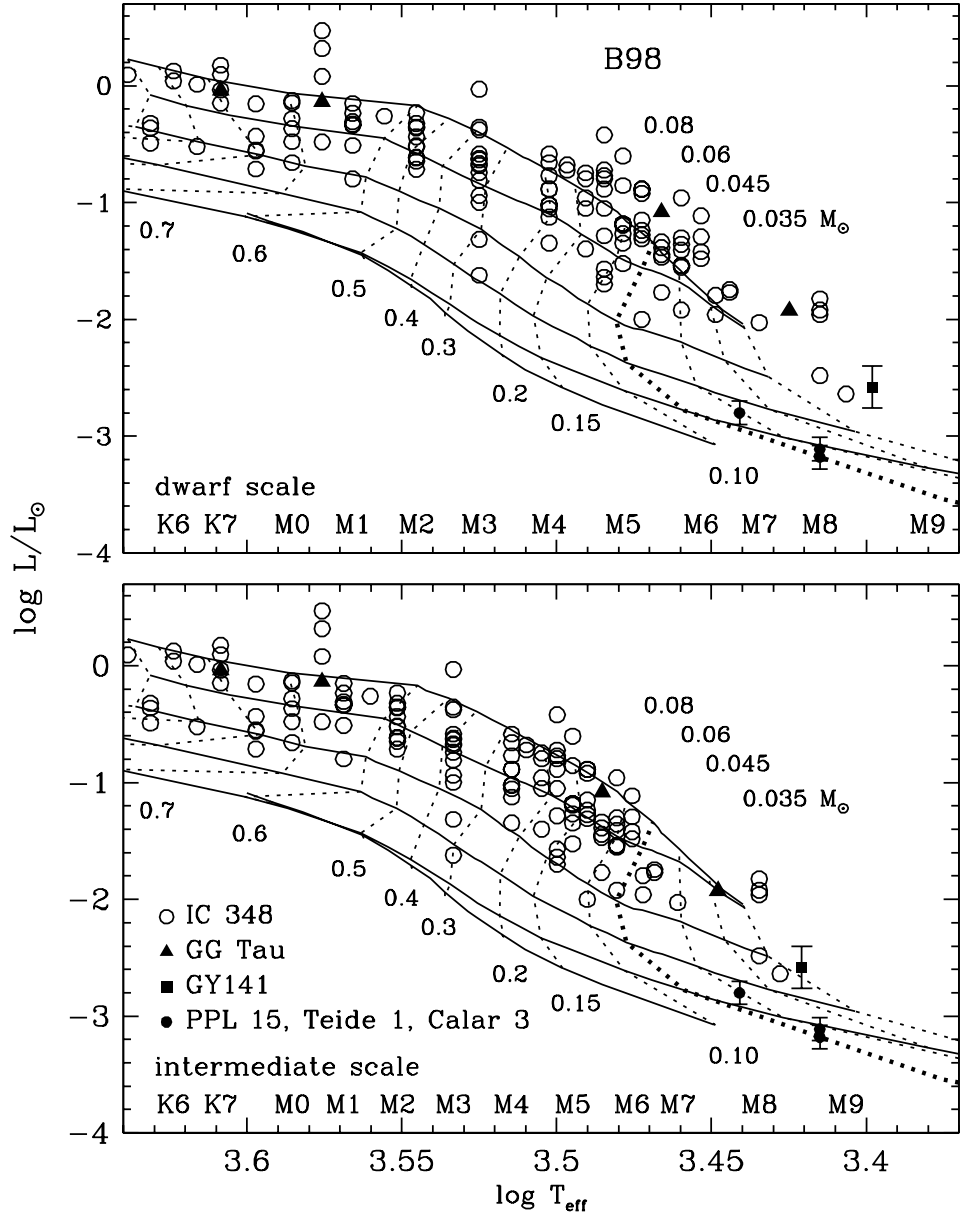}{6.5in}{0}{100}{100}{-306}{-30}
\caption{
H-R diagram for all known late-type sources in IC~348 with the
evolutionary models of B98, where the horizontal
solid lines are isochrones representing ages of 1, 3, 10, 30, and 100~Myr
and the main sequence, from top to bottom.  In the upper panel, 
spectral types have been converted to effective temperatures with a dwarf
temperature scale. In the lower panel, the data are plotted with a
scale such that GG~Tau~Ba and Bb fall on the same model
iscohrone as Aa and Ab. This temperature scale is intermediate between those
of dwarfs and giants, and is given in Table~2 and Figure~\ref{fig:scale}.
The Pleiades brown dwarfs PPL~15, Teide~1, and Calar~3 are plotted with a dwarf
scale. Uncertainties in spectral types are typically $\pm0.25$~subclass. 
}
\label{fig:hr2}
\end{figure}

The data for IC~348 and GG~Tau are shown with the models of B98 ($\alpha=1.9$) 
in Figure~\ref{fig:hr2}. The isochrones imply an age gradient in both 
sets of data, where the less massive objects are progressively younger. 
W99 suggested that a warmer temperature scale for GG~Tau~Ba and Bb could
bring them onto the same model isochrone as Aa and Ab. The necessary departure
from a dwarf scale is increased slightly by the revision of Ba and Bb 
to later spectral types (\S~\ref{sec:compare}). The temperatures of 
Ba and Bb that are required for coevality are shown in Figure~\ref{fig:scale}. 
These temperatures are reasonable for PMS objects since they
are intermediate between dwarf and giant values.  The warmer source Ba is 
closer to the dwarf scale than Bb. As an experiment, this trend is extrapolated
to earlier types until M0, where the dwarf and giant scales converge.
For M8 and M9, the temperatures were selected to be continuous with the values
at earlier types and intermediate between dwarfs and giants with no other 
justification. This intermediate temperature scale is listed in 
Table~2 and illustrated in Figure~\ref{fig:scale}. Using this scale,
the data for GG~Tau and IC~348 are placed on the H-R diagram in the lower panel 
of Figure~\ref{fig:hr2}. As defined, the components of GG~Tau 
are now coeval on the isochrones of B98. In addition,
the locus of IC~348 maintains a constant age and age spread with mass on
these isochrones. Although construction of the intermediate temperature scale
was somewhat ad hoc at types earlier and later than GG~Tau~Ba and Bb, using
this scale with the B98 models is consistent with the constraints
at young ages over a wide range of masses, and should therefore provide the 
relatively reliable ages and masses.

\subsubsection{Implications of Tests}

The theoretical calculations of the evolution of low-mass stars and 
brown dwarfs have become quite sophisticated in recent years. However, 
among the various models there remain large differences in the predicted path
of these objects on the H-R diagram, particularly at ages of $<10$~Myr.
Consequently, the masses and ages estimated for young low-mass objects are
sensitive to the choice of models and the adopted a temperature scale. Given the
observational constraints provided GG~Tau and IC~348, the
models of DM97 are compatible with the dwarf temperature scale, while 
a scale intermediate between those of dwarfs and giants works well with B98. 
The calculations of Burrows et al.\ (1997) do not appear to provide an adequate 
description of young low-mass sources for any reasonable temperature scale.
Both models of B98 and DM97 imply that the hydrogen burning
limit at young ages occurs at a spectral type of $\sim$M6 and that several 
objects in IC~348 fall below the substellar boundary with masses as low as
20-30~$M_J$. 

\section{Conclusion}

I have obtained deep optical photometry and spectroscopy of the young
cluster IC~348 and combined it with previous IR and optical observations.
The conclusions are as follows:

\begin{enumerate}

\item
Expanding on previous optical and IR searches for low-mass stars and brown
dwarfs in IC~348, I have identified a rich population of new low-mass
candidates through $R$ and $I$ photometry. 
Low-resolution optical spectroscopy of a subset of 
these objects has confirmed their youth and late spectral types.

\item
Using the large number of sources in the spectroscopic sample, 
I have described the typical behavior of the optical spectra of young 
late-type objects (M5-M8.5) relative to standard dwarfs and giants.
Overall, averages of dwarf and giant spectra closely resemble the optical
data for young objects and comprise good calibrators of spectral types for
young low-mass populations.

\item
It appears that the intrinsic $R-I$ and $I-J$ colors of young late M objects
are intermediate between the colors of dwarfs and giants, which is 
consistent with the behavior of 
the spectra in these bands. Meanwhile, the intrinsic $J-H$ and $H-K$ 
colors are dwarf-like with an additional $H-K$ excess in
some sources, probably arising from a circumstellar disk or an infalling
envelope. 

\item
After testing the models with empirical isochrones in the form of the
multiple system GG~Tau and the population of IC~348, I find that the 
calculations of Burrows et al.\ (1997) are not consistent with the data 
while the models of DM97 are roughly compatible with the data when a dwarf
temperature scale is used.  The models of B98 produce the best agreement
with observational constraints at young ages, particularly if a temperature 
scale intermediate between those of dwarfs and giants is adopted. 

\item
Under the constraints of the empirical isochrones, both DM97 and B98 
suggest that the hydrogen burning limit occurs near M6 at ages of 
$\lesssim10$~Myr. These models indicate the presence of several new brown 
dwarfs in the spectroscopic sample of IC~348. 

\end{enumerate}

\acknowledgements
I am particularly grateful to the staff of Keck Observatory for their
careful execution of the service observations.  I thank F. Allard, 
I. Baraffe, A. Burrows, N. Calvet, and F. D'Antona for providing their 
most recent calculations and useful advice. Comments on the manuscript 
by L. Hartmann, G. Rieke, J. Stauffer, and R. White are greatly appreciated.
I also thank R. White for access to his spectra of GG~Tau~B. I was funded 
by a postdoctoral fellowship at the Harvard-Smithsonian Center for Astrophysics.
Some of the data presented herein were obtained at 
the W. M. Keck Observatory, which is operated as a scientific partnership 
among the California Institute of Technology, the University of California,
and the National Aeronautics and Space Administration.
The Observatory was made possible by the generous financial support of 
the W. M. Keck Foundation.

\newpage

\clearpage

\clearpage

\clearpage

\clearpage

\clearpage

\clearpage

\clearpage

\clearpage

\clearpage

\begin{deluxetable}{llll}
\tablewidth{0pt}
\tablenum{2}
\tablecaption{Temperature Scales}
\tablehead{\colhead{Spectral} & \multicolumn{3}{c}{$T_{\rm eff}$(K)}\\
\cline{2-4}
\colhead{Type} & \colhead{Dwarf\tablenotemark{a}} & \colhead{Giant\tablenotemark{b}} & \colhead{Intermediate\tablenotemark{c}}
}
\startdata
 M1 & 3680 & 3800 & 3705 \nl
 M2 & 3510 & 3700 & 3560 \nl
 M3 & 3350 & 3590 & 3415 \nl
 M4 & 3180 & 3480 & 3270 \nl
 M5 & 3010 & 3370 & 3125 \nl
 M6 & 2840 & 3250 & 2990 \nl
 M7 & 2720 & 3100 & 2890 \nl
 M8 & 2600 & 2800 & 2720 \nl
 M9 & 2400 & 2650 & 2550 \nl
\enddata
\tablenotetext{a}{Leggett et al.\ 1996 with an extrapolation to types later 
than M6.5.}
\tablenotetext{b}{Perrin et al.\ 1998; Richichi et al.\ 1998; van~Belle
et al.\ 1999.}
\tablenotetext{c}{When this temperature scale is combined with the models
of Baraffe et al.\ 1998, the four components of GG~Tau are coeval.}
\end{deluxetable}

\end{document}